\documentclass[12pt,preprint]{aastex}
 
\slugcomment{To appear in \apj}
 
\shorttitle{The Role of Gas in the Merging of Massive Black Holes in Galactic Nuclei.}
\shortauthors{Escala, Larson, Coppi \& Mardones}

\def\Kpc{\>{\rm Kpc}}

\def\Msun{\>{\rm M_{\odot}}}

\def\msun{M_{\odot}}

\begin{document}
 
\title{The Role of Gas in the Merging of Massive Black Holes in Galactic Nuclei.  I. Black Hole Merging in a Spherical Gas Cloud.} 
\author{Andr\'es Escala}
\affil{
Department of Astronomy, Yale University, New Haven, CT06520-8101, USA \& Departamento de Astronom\'{\i}a, Universidad de Chile, Casilla 36-D, Santiago, Chile.}

\author{Richard B. Larson \& Paolo S. Coppi}
\affil{
Department of Astronomy, Yale University, New Haven, CT06520-8101, USA.}

\author{Diego Mardones}
\affil{
Departamento de Astronom\'{\i}a, Universidad de Chile, Casilla 36-D, Santiago, Chile.}

\begin{abstract}   
Using high-resolution SPH numerical simulations, we investigate the effects of gas on the inspiral and merger of a massive black hole binary. This study is motivated by both observational and theoretical work that indicate the presence of large amounts of gas in the central regions of merging galaxies. N-body simulations have shown that the coalescence of a massive black hole binary eventually stalls in a stellar background. However, our simulations suggest that the massive black hole binary will finally merge if it is embedded in a  gaseous background. Here we present results  in which the gas is assumed to be initially spherical with a relatively smooth distribution. In the early evolution of the binary, the separation decreases due to the gravitational drag exerted by the background gas. In the later stages, when the binary dominates the gravitational potential in its vicinity, the medium responds by forming an ellipsoidal density enhancement whose axis lags behind the binary axis, and this offset produces a torque on the binary that causes continuing loss of angular momentum and is able to reduce the binary separation to distances where gravitational radiation is efficient. Assuming typical parameters from observations of Ultra Luminous Infrared Galaxies, we predict that a black hole binary will merge within $10^{7}$yrs; therefore these results imply that in a merger of gas-rich galaxies, any massive central black holes will coalesce soon after the galaxies merge. Our work thus supports scenarios of massive black hole evolution and growth where hierarchical merging plays an important role. The final coalescence of the black holes leads to gravitational radiation emission that would be detectable out to high redshift by LISA. We show that similar physical effects, which we simulate with higher resolution than previous work, can also be important for the formation of close binary stars.
\end{abstract}  


\section{Introduction}

  Massive black holes (MBHs) are believed to be the power sources for
active galactic nuclei, turning mass into energy through accretion
(Zel'dovich 1964; Salpeter 1964).  The dynamical evidence suggests that every
galaxy with a significant bulge hosts an MBH at its center (Richstone et al.
1998).  Moreover, the data show that the mass of the central black hole is
tightly correlated with the velocity dispersion of the host galaxy's bulge,
yielding the so-called $m_{BH}-\sigma_{c}$ relation (Ferrarese \& Merritt 2000;
Gebhardt et al. 2000).  The estimated masses of the central MBHs in galaxies
range from $10^{6}$ to well above $10^9$~M$_\odot$, making them by far the most
massive known single objects.  However, it remains an open question how
these black holes gain their large masses, and why their masses are strongly
correlated with the dynamical properties of their host galaxies.

   Three possible scenarios have been suggested for massive black hole
formation: (1) They might form by the `monolithic collapse' of a massive gas
cloud to form a single MBH (Rees 1984).  This scenario would require a
mechanism to maintain the gas at a high temperature in order to prevent its
fragmentation into many low-mass objects.  (2) MBHs might form by a series
of mergers of smaller black holes, which might be stellar-mass black holes
or, perhaps more interestingly, `intermediate-mass black holes' (IMBHs)
formed as the remnants of very massive first-generation stars (Bromm, Coppi,
\& Larson 1999, 2002; Abel, Bryan, \& Norman 2002) or by runaway stellar
mergers in very dense environments (Portegies Zwart \& McMillan 2002.)  (3)
MBHs might form by the runaway accretional growth of smaller seed black
holes in the dense nuclear regions of galaxies.  It is also possible that
both gas accretion and the merging of smaller black holes into larger ones,
possibly associated with a succession of galaxy mergers, might be involved
in forming MBHs (Kauffmann \& Haehnelt (2000); Haehnelt 2003; Volonteri, Haardt \& Madau 2003).  In the latter case, it is necessary for black hole mergers to occur on a timescale shorter than that of the galaxy mergers.  It remains unclear, however, just how the central MBHs in galaxies are built up.

   The possibility of black hole mergers in galactic nuclei was first
considered by Begelman, Blandford, \& Rees (1980) in a study of the long-term
evolution of a black hole binary at the center of a dense stellar system.
They showed that the dynamical evolution of such a binary MBH can be divided
into three phases: (a) The two black holes sink toward the center of the
system because of the gravitational drag or `dynamical friction' created by
their interaction with the stellar background (Chandrasekhar 1943; Binney \&
Tremaine 1987).  (b) The orbit of the resulting binary MBH continues to
shrink as 3-body interactions with the surrounding stars continue to extract
energy from the orbit.  (c) If the binary MBH becomes close enough for
gravitational radiation to be important, this becomes an efficient mechanism 
for further angular momentum loss and quickly drives a merger of the two
MBHs.  The occurrence of a merger is however not ensured because 3-body
interactions between the binary MBH and the surrounding stars tend to eject
stars from the central region of the galaxy, depleting the phase-space "loss
cone" of stars that can continue to interact with the binary; this may
prevent a transition between phases (b) and (c), causing the merger to
stall.  More recently this problem has been explored numerically by Makino \&
Ebisuzaki (1996) and Milosavljevic \& Merritt (2001, 2003); these authors
found the coalescence to stall because of the ejection of stars from the
nucleus, although a definitive statement cannot be made because of lack of
sufficient numerical resolution.

   However, this picture does not include the possible role of gas in
driving the evolution of a binary MBH in a galactic nucleus.  Observational
and theoretical work both indicate that large amounts of gas can be present
in the central regions of interacting galaxies, and that this gas can be a
dominant component of these regions.  Numerical simulations show that in a
merger of galaxies containing gas, much of the gas can be driven to the
center by gravitational torques that remove angular momentum from the
shocked gas (Barnes \& Hernquist 1992, 1996; Mihos \& Hernquist 1996); as a
result, more than 60\% of the gas originally present in the merging galaxies
can end up in a massive central concentration or cloud with a diameter of
several hundred parsecs.  Observations of gas-rich interacting galaxies such
as the `Ultraluminous Infrared Galaxies' (ULIRGs) confirm that their central
regions often contain massive and dense clouds of molecular and atomic gas
whose masses are comparable to the total gas content of a large gas-rich
galaxy (Sanders \& Mirabel 1996).  CO data shows that most of the molecular gas  is in relatively smooth interclump medium with dense clumps that account for less than half of the total mass (Downes, Solomon \& Radford 1993; Downes \& Solomon 1998). These massive central gas clouds must have an important influence on the evolution of any central MBH binary that forms
following a galaxy merger.  Because the gas is strongly dissipative, unlike
the stars, it might be expected to remain concentrated near the center and
thus to play a continuing role in driving the evolution of a central binary
MBH.
 
So far, the study of the role gas in driving the evolution of a binary MBH  has  focused on  the interaction of the binary with an accretion disk, where the total gaseous mass is typically smaller than the binary mass. The interaction with an accretion disk can be  effective in driving the orbital decay for high mass ratios ($\rm M_{1} >> M_{2}$; Armitage \& Natarajan 2002; Gould \& Rix 2000). For this mechanism to be effective, a disk mass of at least the same order of magnitude as the secondary MBH is needed. For low mass ratios ($\rm M_{1} \sim  M_{2}$) this mechanism is less effective because strong tidal and/or resonant forces create a circumbinary gap (Lin \& Papaloizou 1979; Goldreich \& Tremaine 1980; Artymowicz \& Lubow 1994,1996). The case of comparable masses is most  relevant for a major galactic merger and the case of our  interest.

In this paper, we study numerically the role of a massive gas cloud, like those seen in the central regions of ULIRGs, in driving the evolution of a binary MBH. We follow the evolution of the binary through many orbits
and close to the point where gravitational radiation becomes important.  As
in the stellar case, the evolution can be divided into three regimes.
First, each black hole is driven closer to the center by the gravitational
drag produced by its interaction with the gaseous background, similar to
the stellar case.  After a central binary MBH has formed it dominates the
dynamics in the region, the orbit of the binary continues to shrink because
of its gravitational interaction with a circumbinary gaseous envelope;
Unlike the stellar case, this envelope remains concentrated around the
binary and is not ejected.  Finally, when the holes come close enough
together, gravitational radiation becomes important and causes rapid merging
of the binary.  Here we present results for a relatively simple idealized
case in which the gas is assumed to be in a spherical cloud supported by a
high virial temperature, so that the gas retains a nearly spherical and
relatively smooth distribution. Our model studies the effect of the relatively smooth medium that accounts for most of the molecular mass in the center of a typical ULIRG (Downes, Solomon \& Radford 1993; Downes \& Solomon 1998). Subsequent papers will present results of calculations in progress for  cases in which the gas is cooler and has a more flattened or clumpy spatial distribution with rotation.

We start our analysis validating the use of the code in \S 2 by
performing several tests. In \S 3 we perform a numerical study of the
differences between the dynamical friction  in a gaseous background
and the dynamical friction in the stellar case. In \S 4 we study the evolution
of a binary MBH in a gaseous sphere. Finally in \S 5 we apply our
results to the inner regions of a ULIRG, showing that the merging of two massive black holes in such a system can occur in a time as short as $10^{7}$ years.

\section{Description of the Code}

We use the  Smooth Particle Hydrodynamics (SPH) code called GADGET (Springel, Yoshida \& White 2001) to study the dynamical evolution of a black
hole binary in a gaseous medium.  GADGET has been successfully tested
in a broad
range of problems such as galaxy interactions and collisions, cosmological
studies of large scale structure and  high resolution simulations of the formation of galaxy clusters (Springel, Yoshida \& White 2001).  The code has many special
features such as shear-reduced artificial viscosity, efficient
cell-opening criteria, individual time-steps of arbitrary size for all particles, good scalability
and an efficient integrator in the linear regime of gravitational
clustering. All of these features make GADGET  suitable for our work.

The SPH code allows one to study both  collisional and collisionless systems.  Stars and 
MBHs are modeled as a self-gravitating collisionless fluid  governed by the 
Vlasov Equation (Binney \& Tremaine 1987, eq 4-13c) 
\begin{equation}
\frac{df}{dt} = \frac{\partial f}{\partial t} + {\bf v} \frac{\partial f}{\partial {\bf r}} - \frac{\partial \Phi}{\partial {\bf r}} \, \frac{\partial f}{\partial {\bf v}} = 0,
\end{equation}
where the gravitational potential $\Phi$  satisfies Poisson's equation
\begin{equation}
\nabla^{2}\Phi({\bf r},t) = 4\pi G \, \int f({\bf r},{\bf v},t) d{\bf v}, 
\end{equation}
and $f({\bf r},{\bf v},t) d{\bf v}$ 
is the phase space mass density. It is very difficult to solve this set of coupled differential equations directly. Instead 
we use the N-body method, a montecarlo approach where the  fluid is
represented by N particles. The accuracy in this method depends on  the number of particles used.

The gaseous medium is modeled as an ideal, viscid gas, governed by the
standard laws of fluid dynamics: 
the equation of continuity
\begin{equation}
\frac{d\rho}{dt} + \rho\nabla\cdot{\bf v}= 0,
\end{equation}
and the momentum equation 
\begin{equation} 
\frac{d{\bf v}}{dt} = -\frac{{\bf \nabla}P}{\rho} \, + \, {\bf a^{visc}} \, - \, {\bf \nabla}\Phi,
\end{equation}
where the time derivatives are Lagrangian $(d/dt = \partial/\partial t \, +
\, {\bf v}\cdot\nabla)$. To close the set of equations, we use a polytropic equation of state
\begin{equation}
p = K \, \rho^{\gamma},
\end{equation}
with polytropic index $\gamma$ equal to 1 (isothermal) or 5/3 (adiabatic). We   solve the equations of fluid dynamics  by  the  Lagrangian  technique  called SPH (see Monaghan 1992 for a review). In SPH the fluid elements  constituting the system are sampled and represented by particles. By design, the SPH technique  is   free of geometrical restrictions and the resolution is automatically increased in the denser regions, because  particles are  concentrated in those regions.

In order to validate the proper use of GADGET, we did  several
additional tests to those   performed by the authors of the code (Springel,
Yoshida \& White 2001); e.g. the  pulsation of an adiabatic sphere, the
Evrard collapse (Evrard 1988), and the collapse of a primordial rotating cloud with  cooling, all of which yielded with satisfactory results. A full description of the implementation of these tests is given in Appendix A. 



\section{Dynamical Friction in a Gaseous Medium}
Dynamical friction is the gravitational drag force that  a massive
body suffers while moving through a collisionless background medium. It has been widely studied in
the literature since the concept was introduced  by Chandrasekhar in
1943. He computed the deceleration that a massive body suffers as it
moves through an infinite homogeneous  medium of smaller particles
that exchange momentum with the body, ignoring the self-gravity of
the medium. Because Chandrasekhar assumed an  infinite homogeneous background medium,  the formula diverges. An heuristic device was introduced to remove the divergence, the maximum impact parameter $({\rm b}_{\rm max})$, which is related to the size of the system. The combination of this parameter with another  that can be interpreted dimensionally as a minimum impact parameter appears in Chandrasekhar's formula as a Coulomb logarithm (ln(${\rm b}_{\rm max}$/${\rm b}_{\rm min}$) = ln$\Lambda$). 

Despite the heuristic treatment of the Coulomb logarithm, numerical simulations have shown that Chandrasekhar's formula  provides a remarkably accurate description of the drag that   a massive body orbiting in a stellar system experiences(Lin \& Tremaine 1983; Bontekoe \& van Albada 1987; Cora et al. 1997). The procedure used to  compare  Chandrasekhar's formula with simulations is done by  evaluating the formula using  local quantities at the instantaneous position of the body, leaving ln$\Lambda$ as a parameter that can be adjusted to achieve the best fit.

Less effort has been devoted to the study of the drag force experienced by
  a gravitating body  moving through a gaseous background
medium.  Ostriker (1999) did an  analytical
study of the gravitational drag that  a massive body suffers while
traveling through a infinite uniform gaseous medium. The drag force
arises from the gravitational attraction between the perturber and the
density wake behind the body. The force is computed
using time-dependent linear perturbation theory. Ostriker's main
result was that the gaseous drag is enhanced for supersonic
motions (Mach numbers of the gravitating body in the range of 1--2.5),
compared to Chandrasekhar's formula. Additionally, Ostriker found a  non-vanishing gravitational drag for subsonic motions, although less efficient than the stellar case. Previous work (Rephaeli \& Salpeter 1980) based on a steady state theory had found that the gravitational drag vanishes for subsonic motions due to the front-back symmetry of the perturbed density distribution.

Several groups have  numerically computed the gravitational drag force that  a massive body suffers while traveling through a infinite uniform gaseous medium, as a by-product of a study of the Bondi-Hoyle accretion problem (see Shima et al. 1985, Ruffert 1996, and references
therein), and they found results consistent with  Chandrasekhar's formula. Sanchez-Salcedo \& Brandenburg (2001) extensively 
numerically tested  the validity of Ostriker's formula, in
a study of orbital decay analogous to  the problem of orbital decay in spherical stellar system where Chandrasekhar's formula is a good approximation (Lin \& Tremaine
1983; Bontekoe \& van Albada 1987; Cora et al. 1997). They  found that Ostriker's formula is able to explain successfully the evolution of the perturber when the motion is highly supersonic, although this theory fails if the motion is only  slightly supersonic. They also found that the dynamical friction time-scale is a factor  of two smaller in the case of a gaseous sphere than  in the corresponding stellar system.

In this paper we model the evolution of a binary MBH  in a gaseous
sphere. This is basically a gravitational drag  problem, as long
as the evolution of the binary is determined by the gravitational
interaction of each black hole with its local surrounding medium. During the later stages, when the binary MBH dominates the gravitational potential around the orbit,  the evolution of the binary is determined by the interaction of a binary MBH with a circumbinary gaseous envelope; we call this regime a "hard binary" in  analogy to the stellar
dynamics problem (Quinlan 1996, Milosvljevic \& Merritt 2001). For both stellar and gaseous backgrounds, the dynamical friction approximation is no longer valid when the MBH's are at distances where $M_{\rm binary} \sim M_{\rm (star,gas)}(r<r_{\rm binary})$.   

\subsection{Model Setup}

In our model we assume that the background consists of a self-gravitating gas sphere in hydrostatic
equilibrium, within which a massive particle is initially orbiting in a circular orbit. The equilibrium density profile is the  nonsingular isothermal sphere (NSIS)
\begin{equation}
\rho(r) = \rho_{c} \, e^{-\Phi(r)/\Phi_{c}}, 
\label{isothdens}
\end{equation}
where $\Phi(r)/\Phi_{c}$ is the gravitational potential  obtained by solving
 the isothermal Lane-Emden equation numerically. A Bonnor-Ebert stability
analysis was carried out to guarantee that the NSIS is truncated at a radius where the equilibrium is  stable. The SPH particles are distributed according the density profile given by Eq $\ref{isothdens}$, by employing a standard comparison rejection method (Press et al. 1992). We relax the system using the technique developed by Lucy (1977) in order to reduce the numerical noise. The units used in this paper are such $M^{\rm gas}_{\rm total}$ = 2, $C_{\rm S}$ = 2.5 and the sound crossing time is 1. 






We check that our calculations converge as we increase the
resolution of the code, running the same problem with different numbers of
SPH particles (25000, 50000, 100000, 200000, and 400000).  On each run
we introduce a massive collisionless particle, the MBH, in the
same initial circular orbit and follow its evolution through several orbits.
The mass of the MBH is  1\% of the total gas mass.

\begin{figure}
\plotone{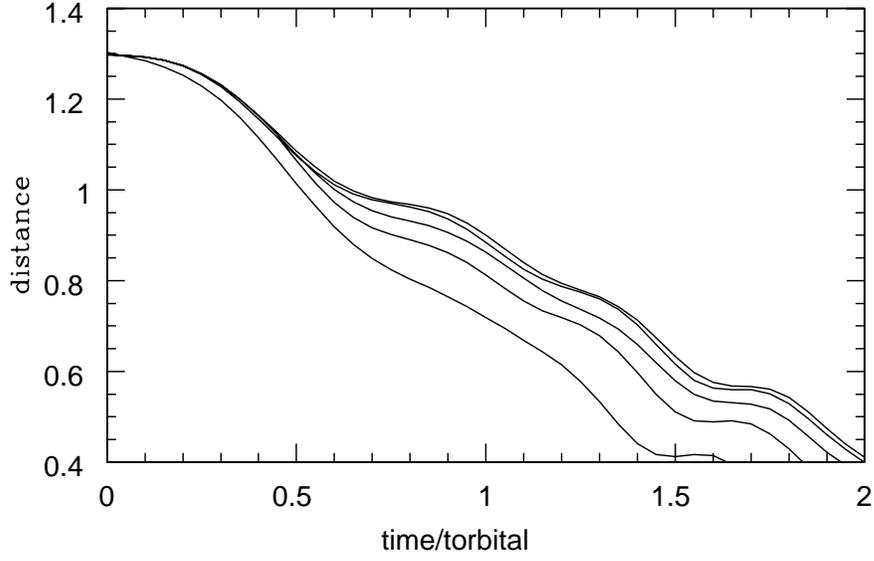}
\caption{Convergence Study. Evolution of the distance of a single MBH from the center of mass of the gaseous sphere. The solid lines (from top to bottom) represent different runs with N=400,000 200,000 100,000 50,000 25,000 respectively.  
\label{fig5}}
\end{figure}

Figure $\ref{fig5}$ shows the evolution of the radial distance
of the MBH from the center of mass of the gaseous system. The curves are the 
results for the different resolution levels, the upper line
has 400,000  particles and lower lines indicate the results for
runs with decreasing number of particles.  The solution rapidly converges
as we increase the number of particles, and has an adequate level of
accuracy with 100,000 particles or more.

\begin{figure}
\plotone{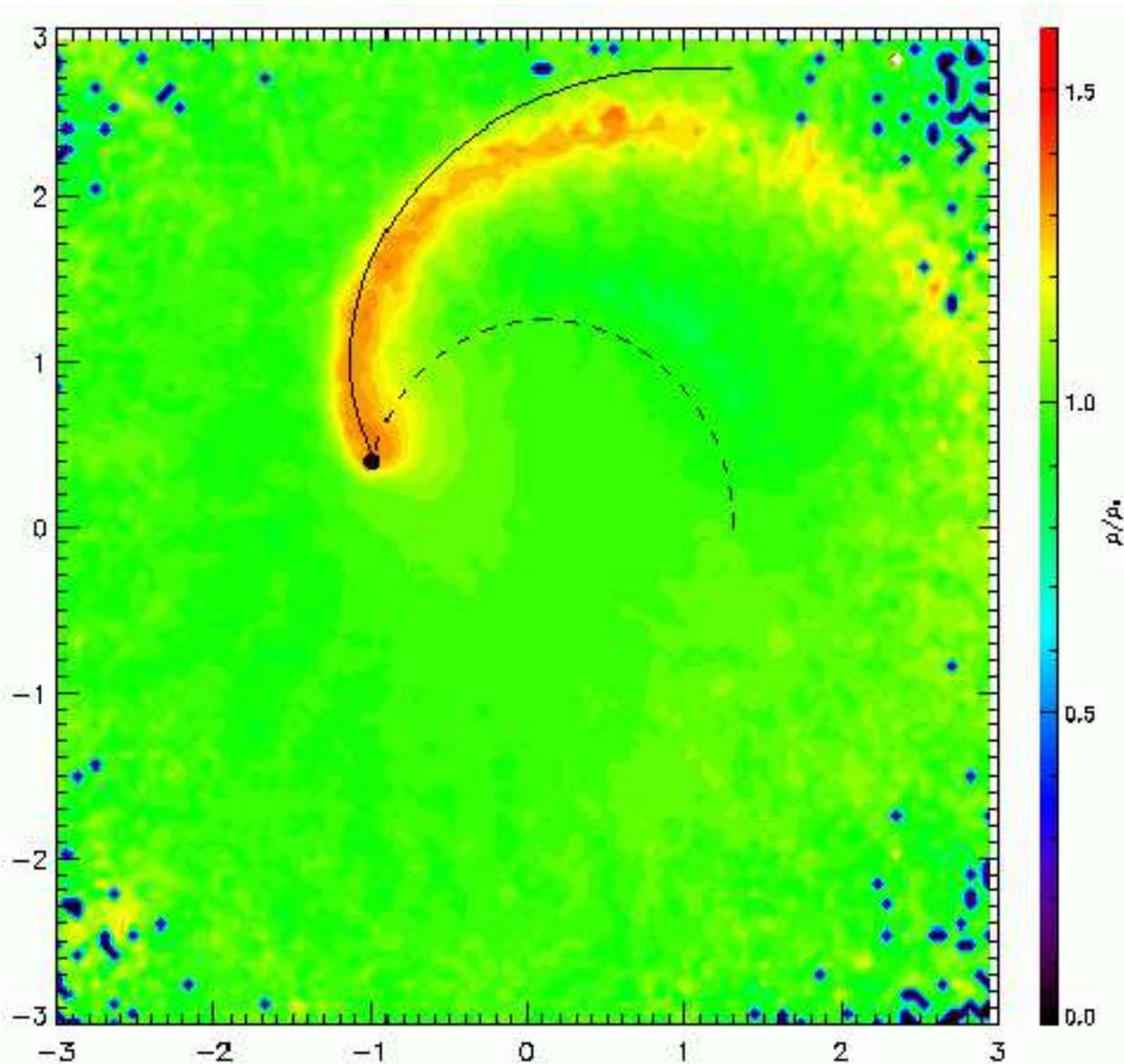}
\caption{Spiral Shock produced by an orbiting MBH. The color gradient represents the density
  enhancement  $(\rho(t=1)/\rho(t=0))$  produced by a single MBH in the
  gaseous isothermal medium in the plane of the orbit (z=0). The dashed black  
  line is the orbit of the MBH between t=0 and t=1. The black dot  indicates the position of the MBH at t=1. The solid black curve is the shock locus predicted by the simple approximation discussed in section 3.1. 
\label{fig5.5}}
\end{figure}

The interaction between the massive body and the gas produces a density enhancement in
the gas sphere as shown in Fig. $\ref{fig5.5}$.  The colors represent the density 
enhancement $[\rho(r,t)/\rho(r,0)]$ in the plane of the orbit (z=0) for ${\rm t}=1$.  The dashed black curve is
the orbit of the MBH between ${\rm t}=0$ and ${\rm t}=1$.

The initial circular velocity of the massive body is that predicted for an isothermal sphere: $v_{\rm BH} \, = \, \sqrt 2 \, C_{\rm S}$. The supersonic motion 
produces a spiral shock that propagates outwards with a mildly supersonic
speed $(1.1 \, C_{\rm S})$.  A qualitative explanation for the spiral form of the shock is  that the MBH accelerates the gas by the reaction to
the gravitational drag force. The direction of the resulting impulse is tangential
to the orbit and it is assumed that the disturbance propagates in a straight line with a speed of $1.1 \, C_{\rm S}$. The solid black line  shows the locus of the propagated disturbances, showing a good
agreement with the spiral density enhancement.  This feature is qualitatively
the same as that found by  Sanchez-Salcedo \& Brandenburg (2001)  in a similar
analysis (see their Fig. 1). However, we have calculated that 90\% of the gravitational drag comes from the region within a radius of 0.3 measured from the MBH; therefore the details of the spiral shock do not affect the motion of the MBH.

\subsection{Comparison of the Dynamical Friction in a Stellar vs a Gaseous Medium}
 
We want to  compare the dynamical friction in a gaseous background with that in  
a  stellar medium. The dynamical friction can
be expressed as the product of two factors: a factor with the
dimensions of force, which is the same in the stellar and gaseous
cases, and a dimensionless factor, $f^{\rm (star,gas)}$, which depends on the nature of the background medium:
\begin{equation}
F_{\rm DF}^{(star,gas)} = -4\pi\rho\left(\frac{G M_{\rm BH}}{C_{\rm S}}\right)^{2}  
                       \times f^{\rm (star,gas)}\left({\cal M} \right)
\label{fdf}
\end{equation}
where ${\cal M}$=$v_{\rm BH}$/$C_{\rm S}$. In this context, $C_{\rm S}$ is
 the sound speed for the gaseous background or the velocity dispersion for the
  stellar background. In the stellar case, the expression $f^{\rm (star)}$  is the following (Chadrasekhar 1943) 
\begin{equation}
f^{\rm (star)} = {\rm ln}\left(\frac{r_{\rm max}}{r_{\rm min}}\right) \, {\cal M}^{-2} \, 
\left[{\rm erf}\left(\frac{{\cal M}}{\sqrt{2}}\right) - \sqrt{\frac{2}{\pi}} {\cal M}  e^{-\frac{{\cal M}^{2}}{2}}\right] \, ;
\label{chandra}
\end{equation}
similarly, Ostriker (1999) computed $f^{\rm (gas)}$ for both supersonic and subsonic perturbers
\begin{equation}
f^{\rm (gas)}_{\rm subsonic} = \frac{1}{2{\cal M}^{2}} \,  {\rm ln} \left(\frac{1+{\cal M}}{1-{\cal M}}\right) - \frac{1}{{\cal M}} \, ,
\label{ostrikersub}
\end{equation}
\begin{equation}
f^{\rm (gas)}_{\rm supersonic} =  \frac{1}{{\cal M}^{2}} \, \left[ 0.5 \cdot {\rm ln} \left(\frac{{\cal M}+1}{{\cal M}-1}\right) -
{\rm ln}\left(\frac{r_{\rm max}}{r_{\rm min}}\right) \right] \, .
\label{ostrikersup}
\end{equation}
In order to  numerically evaluate the   differences, we set up models with exactly the same dimensional parameters $(M_{\rm BH}, \, \rho \,\, and \,\, v_{\rm BH}={\cal M} C_{\rm S})$, for the
gaseous and stellar mediums.  To do this we start with self-gravitating
isothermal spheres that are  in hydrostatic equilibrium (for a gaseous
system) and in  dynamical equilibrium (for a stellar system; \S 4.4.3b of Binney \&
Tremaine 1987). The goal is to use  these models to compare directly the gravitational drag in a
collisionless medium with that in a collisional one.

To investigate the difference between the gaseous and stellar cases, we follow the evolution of a single MBH (with 1\% of the total gaseous/stellar mass) in the same initial circular orbit, in both  the gaseous and stellar equilibrium configurations. The velocity of the MBH is initially supersonic $(v_{\rm BH}\sim 1.4 \, C_{\rm S})$ and remains barely supersonic through most of the 
simulation. We initialize both runs with the same value for $-4\pi \rho \frac{(G
M_{\rm BH})^{2} }{C_{\rm S}^{2}}$, thus, any difference in the evolution 
reflects purely the effects of the term $f^{\rm (star,gas)}$.  Each of these runs has a
resolution of 100,000 stars/gas particles.

\begin{figure}
\plotone{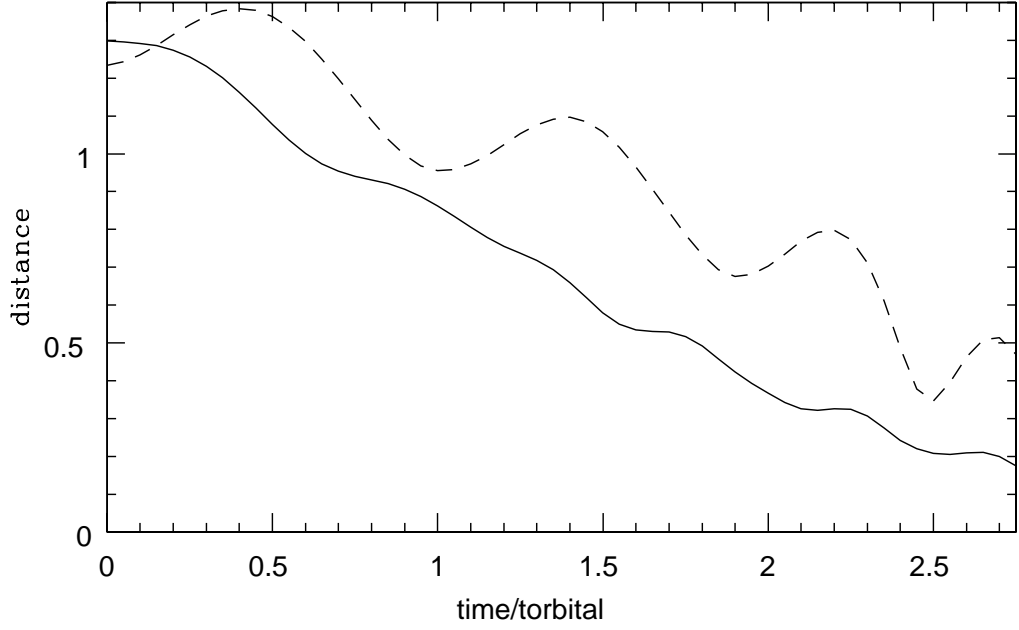}
\caption{Comparison between the Stellar and Gaseous Cases. The plot shows the evolution of the distance  of the MBH measured from  the center of mass of the isothermal sphere. The lines are the predictions for the gaseous sphere (continuous) and the stellar sphere (dashed).
\label{fig6}}
\end{figure}

\begin{figure}
\plotone{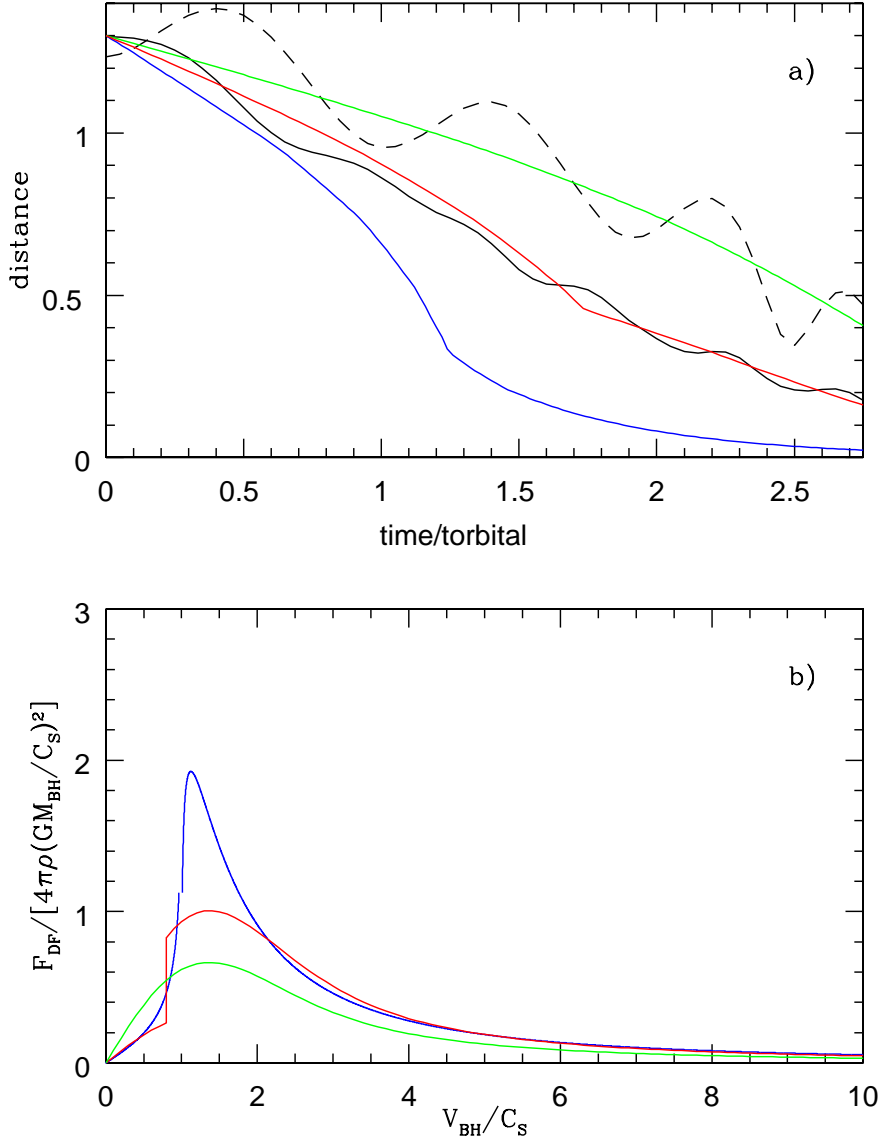}
\caption{a: The plot shows the evolution of the distance (in code
  units) of the MBH to   the center of mass of the
  isothermal sphere. The lines are the predictions for: gaseous sphere
  (black continuous), stellar sphere (black dashed). The green line is
  Chandrasekhar's  formula (for ${\rm ln}\Lambda$ = 3.1). The blue line 
is the prediction
  of  Ostriker's formula for the same  ${\rm ln}\Lambda$ = 3.1, and the red
  line is the prediction
  using the parametric function shown in red in the part b) of the
  figure. b: The dimensionless factor $f^{\rm (star,gas)}$ of the dynamical
  friction force as a function of $v_{\rm BH}/C_{\rm S}$, used in different
  fits in the top figure. The colors match those in Fig. 
$\ref{fig7b}$a. The initial Mach number of the BH is $\sqrt 2$.
\label{fig7b}}
\end{figure}

Fig. $\ref{fig6}$ shows the evolution of the radial distance of the MBH from the center of mass of the system, represented by a dashed line in the stellar case and by the continuous
line in the  gaseous medium.  The distance is plotted in
code units and  time in units of the initial orbital period. The distance of the MBH from the center changes
from 1.4 at t=0 to  0.6 at t=1.5 ${\rm t}_{\rm orbital}$ in the gaseous
case. However, in the stellar case the average distance is not 0.6 until t=2.3 ${\rm t}_{\rm orbital}$. This suggests that the drag timescale is approximately a factor of 1.5 shorter for the gas
configuration. The velocity of the BH remains  supersonic through
most of the  run,  thus we are measuring $f^{\rm (star,gas)}$ close to 
its peak value (see Fig $\ref{fig7b}$b).

An attempt to explain  the behavior of the MBH (for both mediums, in
terms of Eqs $\ref{chandra}$-$\ref{ostrikersup}$) is shown in  Fig. $\ref{fig7b}$a. The dashed and continuous black lines represent  the radial distance of the MBH in the stellar  and gaseous medium. The green line is the fit of the Chandrasekhar formula (Eq $\ref{chandra}$) to the evolution of the MBH in the stellar case, with ${\rm ln}({\rm b}_{\rm max}/{\rm b}_{\rm min})$=${\rm ln}\Lambda$=3.1 as the best fit value for the coulomb logarithm. This value is consistent with results found by other authors for orbiting objects with  masses of the order of  1\% of the total mass in stars (Lin \& Tremaine 1983, Cora et al. 1997).

We try to reproduce the gaseous simulation using  Ostriker's
formula (Eqs $\ref{ostrikersub}$ \& $\ref{ostrikersup}$) for
${\rm ln}\Lambda$=3.1, and the result is shown by the blue line in
Fig. $\ref{fig7b}$a. It clearly  overestimates our prediction for the
gaseous drag by a factor of 1.5 (or 2.3 relative to the stellar
case). This result is not surprising, because Sanchez-Salcedo \&
Brandenburg (2001) had already found that  Ostriker's formula overestimates in
 predicting the gaseous gravitational drag  for a
body  moving at transonic velocities ($v_{\rm BH}=\sqrt2 C_{\rm S}$ at the
start of our run). Fig. $\ref{fig7b}$b shows the dimensionless factor of the dynamical friction force as a function of $v_{\rm BH}/C_{\rm S}$, as used in different fits in Fig. $\ref{fig7b}$a. In particular Ostriker's formula is  represented  by  the blue line in Fig. $\ref{fig7b}$b.

Due to the failure of  Ostriker's formula, we consider   a parametric function for $f^{\rm (gas)}({\cal M})$ as a better approximation. This function has the same form of the Eq $\ref{chandra}$ but with two choices for ${\rm ln}\Lambda$, with values of 4.7 for ${\cal M}$ $\geq$ 0.8 and 1.5 ${\cal M}$ $<$ 0.8. The red curve in Fig. $\ref{fig7b}$b is a plot of this function and is qualitatively like Ostriker's formula but less peaked.  The result obtained with this parametric function is considerably better, showing good agreement with our SPH run as seen in Fig. $\ref{fig7b}$a.

\section{Evolution of a binary MBH}

In \S 3 we looked at the effect of the gravitational drag on the evolution of a single MBH in a gaseous sphere. In this section, we  study  the effect of  gravitational drag on the evolution of a binary MBH in a gaseous sphere, with   the aim of   determining the timescales  for hardening of the binary and possible merging.  To investigate this problem, we follow the evolution of two massive bodies of equal mass in the same stable gas configuration as in  \S 3. As in the single MBH case, the gas mass is much greater
than the mass in the binary and the gas mass  within the binary  produces the acceleration
necessary to maintain the circular orbit. As long as the condition  $M_{\rm gas}(r<r_{\rm binary}) \geq M_{\rm binary}$ is satisfied, the evolution of the binary will be determined by the gravitational interaction of each black hole with its surrounding medium, as in the problem studied in $\S$ 3.3. We use a resolution of 100,000 SPH particles to model the gas sphere and  run the code for various  MBH masses.

\begin{figure}
\plotone{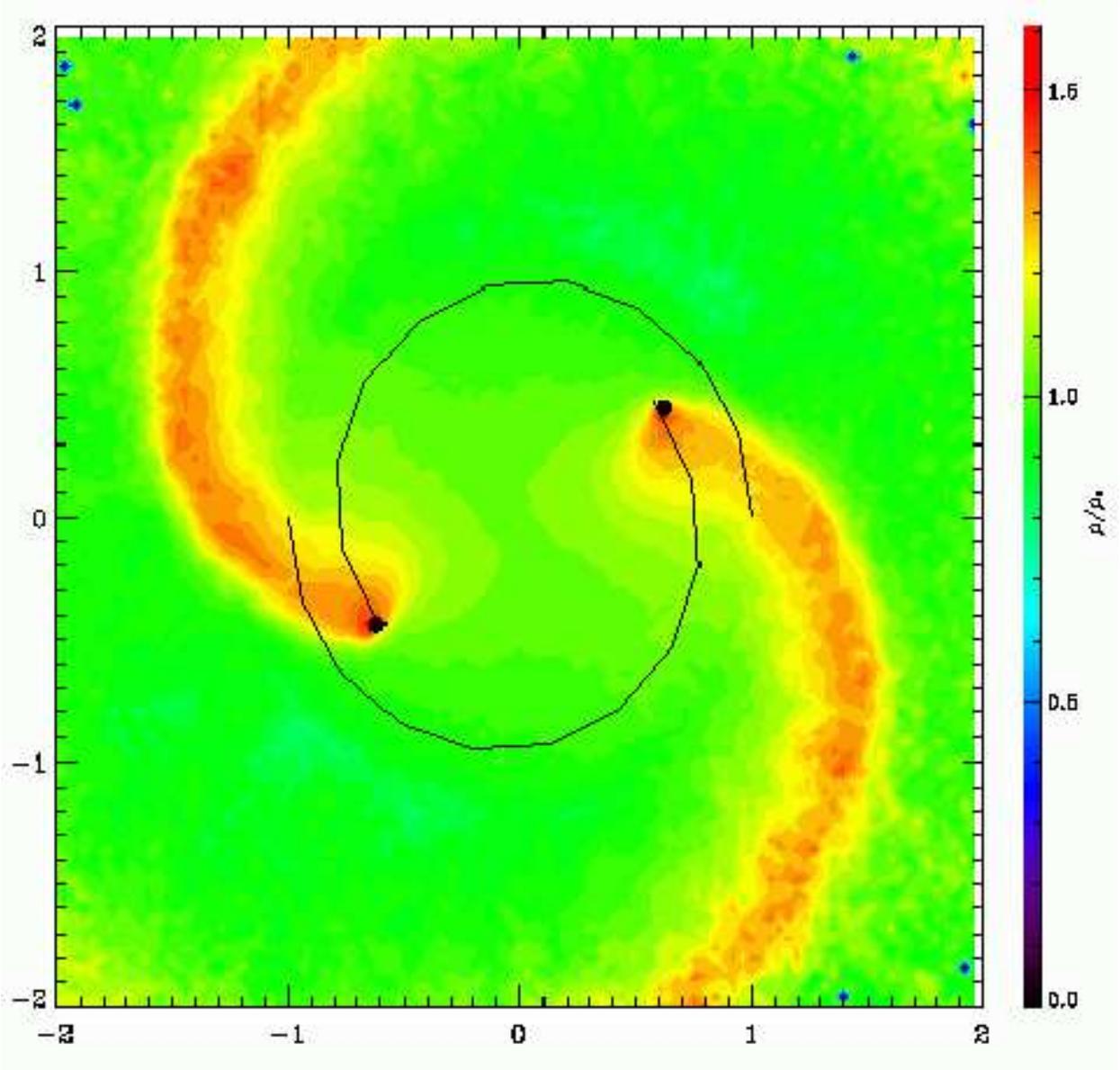}
\caption{Density Enhancement in the Isothermal Medium. The colors in the figure represent the density enhancement $(\rho(t)/\rho(0))$ in the isothermal sphere  in the plane of the orbit (z=0) for  t=1. The black curves are the individual   orbits of the MBH's between t=0 and t=1. In this simulation the mass of  the binary is  2\% of the mass of the gas. 
\label{fig8}}
\end{figure}

Fig. $\ref{fig8}$ shows the response of the gaseous medium to the presence of
the MBH binary. The colors in the figure represent the density enhancement
$(\rho(r,t)/\rho(r,0))$ in the plane of the orbit of the binary.  A couple of
spiral shocks propagate outwards, one caused  by each MBH. The black curves show the individual orbits of the MBHs between t=0 and t=1. This figure shows the simulation for the case of a MBH
with 1\%  of the total mass of the gas sphere. These density
enhancements appear similar to that formed by a single MBH with the
same mass (see Fig. $\ref{fig5.5}$).

\begin{figure}
\plotone{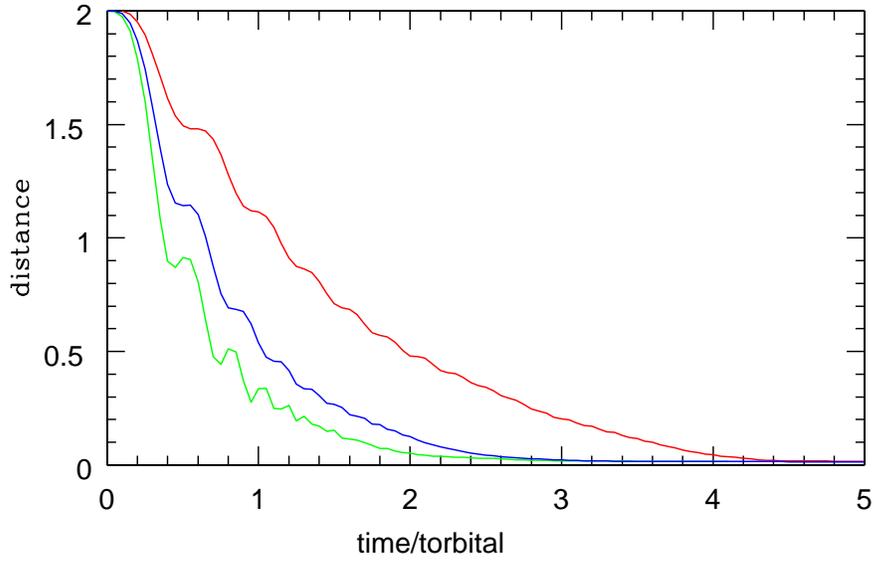}
\caption{The plot  shows the evolution of the
  binary separation in the isothermal sphere, and the different curves represent the separation  between the black holes for different mass  ratios: $M_{\rm binary}=0.02M_{\rm gas}$ (red), $M_{\rm binary} = 0.06M_{\rm gas}$ (blue), $M_{\rm binary} = 0.1M_{\rm gas}$ (green).
\label{fig10}}
\end{figure}

In our study we consider three cases with different black hole masses of 1\%, 3\% and 5\% of the total gas mass. Figure $\ref{fig10}$ shows the evolution of the binary separation in these three cases over
several orbits. The different curves
represent the binary separation for different binary to gas mass ratios:
$M_{\rm binary}=0.02M_{\rm gas}$ (red), $M_{\rm binary} = 0.06M_{\rm gas}$ (blue),
$M_{\rm binary} = 0.1M_{\rm gas}$ (green).  The binary evolution has a strong
dependence on the MBH masses, the more massive cases suffering a
stronger  deceleration. The time required for the binary separation to decrease by a factor of 2 decreases from  ${\rm t} \sim 1.25$ for  $M_{\rm binary}=0.02M_{\rm gas}$  to  ${\rm t} \sim 0.5$ for $M_{\rm binary} = 0.1M_{\rm gas}$ (times  given in units of the  initial orbital period). 

Fig. $\ref{fig10}$ shows clearly  the fast decay
of the binary separation. For all mass ratios, the binary MBH reaches
the resolution limit, the gravitational softening length, in only 2 to 4
initial orbital periods, depending on the binary mass. These
timescales are shorter than in the stellar case   because, as seen in \S 3.3, the 
timescales are a factor 1.5 faster
due to the collisional nature of the background medium. 


At the final separation, gravitational radiation is still an inefficient mechanism for angular momentum loss, but these results are limited by the gravitational softening length. Simulation of later stages of evolution therefore requires higher numerical resolution, as is described below.

\subsection{Further Evolution Using Higher Resolution Simulation}

To continue the evolution of the binary it is necessary to reduce the softening length, something that is extremely expensive computationally. We choose to reduce it from $\epsilon_{\rm soft}$=0.01  to 0.001 in order to follow the evolution of the binary separation by one more order of magnitude. We also  need to increase the resolution to  resolve the region inside r = 0.1  where 
 most of the dynamics occurs. We use  the particle splitting procedure described by Bromm (2000), Kitsionas (2000), and in Appendix B to achieve this goal.
The  procedure is applied to the case  where  each MBH has 1\% of the total gas mass in gas. We split the SPH particles at a  time of 3 initial orbital periods when the binary enters  the r $<$ 0.1 region; for  each parent particle we introduce  $N_{\rm split}$=20 child particles and therefore we increase the number of particles to 2,000,000.

\begin{figure}
\plotone{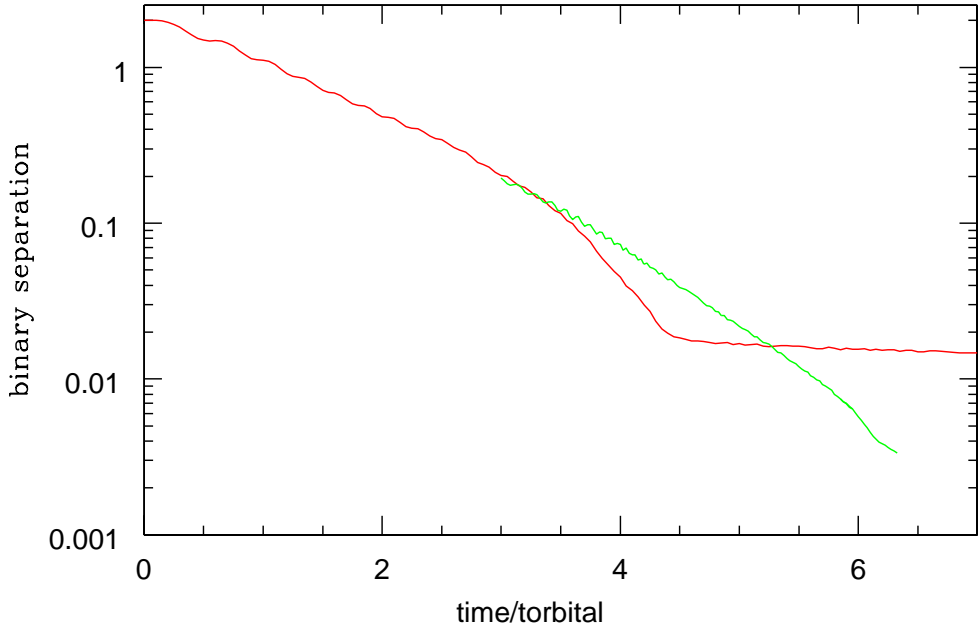}
\caption{The plot shows  the complete evolution of the binary
separation. The red curve is the same coarse
calculation shown in Fig. $\ref{fig10}$, for $M_{\rm binary}=0.02M_{\rm gas}
\,$, but  on a logarithmic scale. The green curve is the separation of
the MBHs  in the high resolution (fine) calculation.
\label{fig9b}}
\end{figure}

Fig. $\ref{fig9b}$ shows the  evolution of the binary, from the beginning  of the low-resolution simulation to the end of the   high-resolution simulation. The red curve is the same  calculation shown in Fig. $\ref{fig10}$  but  on a logarithmic scale, and the green curve shows the MBH's separation in the high-resolution calculation. The green curve shows an almost linear trend in  logarithmic separation that differs considerably from the behavior shown by the low-resolution simulation. At the end of this high resolution simulation, the binary separation has decreased by almost three orders of magnitude. 

\begin{figure}
\plotone{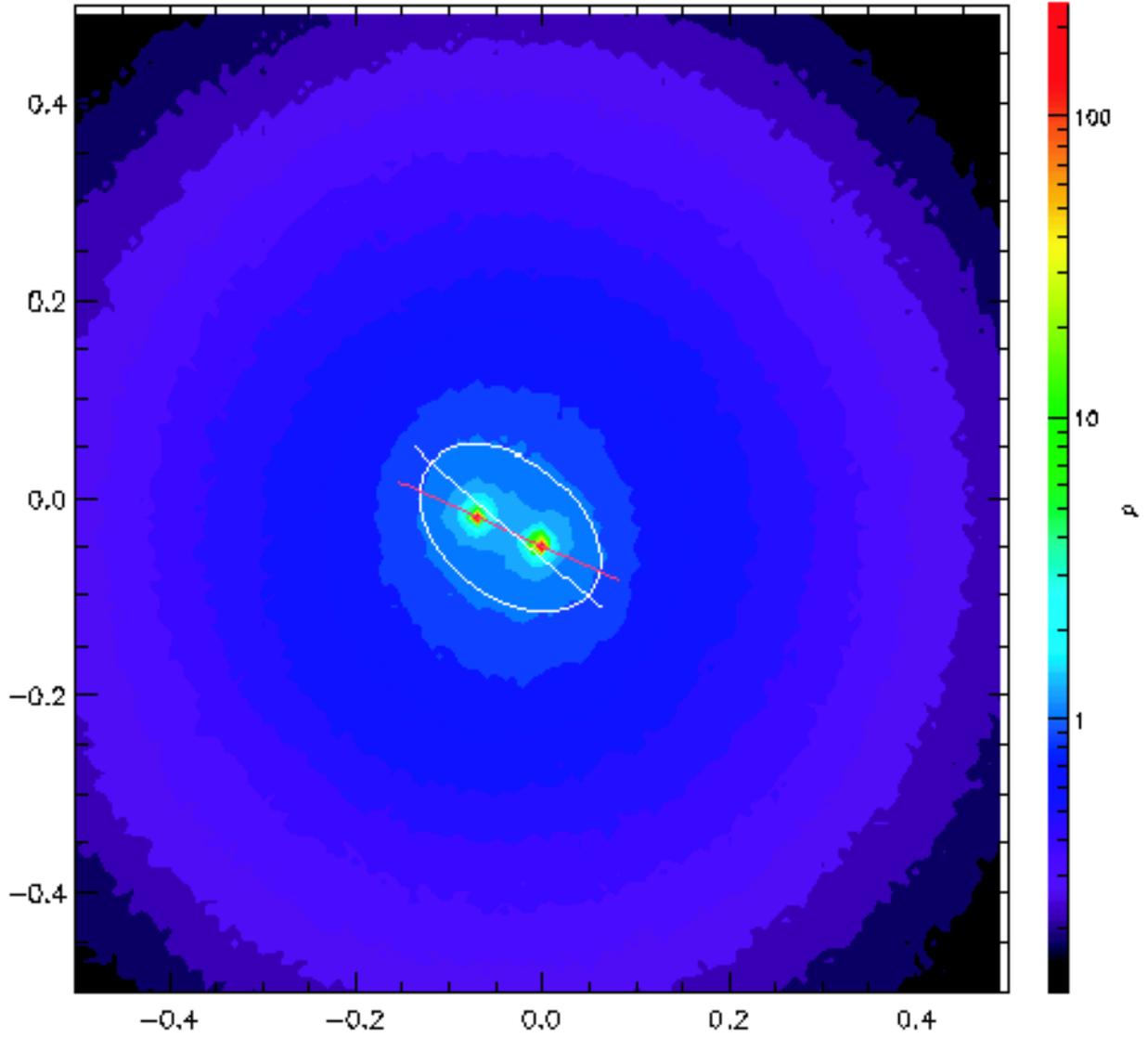}  
\caption{Density, in  the plane of the orbit (z=0) for  $\rm t=4t_{orbital}$. The
  color scale is logarithmic. The
  two density peaks, in red, coincide with the location of the
  MBHs. The density in the inner regions $(r \leq 0.1)$  has  an
  ellipsoidal shape, with an axis inclined with respect to the binary axis.  
\label{fig11b}}
\end{figure}

The overall evolution of the  binary MBH can be divided into two physical regimes: i)Up to t=3 ${\rm t}_{\rm orbital}$ where  $M_{\rm bin} \leq M_{\rm gas}(r<r_{\rm bin})$, the  angular momentum loss of the binary is governed by the gravitational drag exerted by the gaseous medium. ii) After t=4 ${\rm t}_{\rm orbital}$  the binary satisfies the condition  $(M_{\rm bin} >> M_{\rm gas}(r<r_{\rm bin}))$, therefore the binary completely dominates  the gravitational potential in its vicinity and  the response of the  medium to that gravitational field is  an ellipsoidal density distribution as seen in Fig. $\ref{fig11b}$. The axis of the ellipsoid is not coincident with the binary axis but lags behind it, and this  offset  produces a torque on the binary that is now  responsible for the angular momentum loss. The evolution of the system is approximately self similar because the size of the ellipsoid depends on the binary separation and this is responsible for the nearly  exponential decay after t=4 ${\rm t}_{\rm orbital}$  (Fig. $\ref{fig9b}$). In the next section we  study this torque in more detail. From  t=3 ${\rm t}_{\rm orbital}$ to 4 ${\rm t}_{\rm orbital}$ is a transition period between these two  physical regimes.

\subsection{Ellipsoidal Torque}

Our numerical results for the final stage of evolution can be understood in terms of a simple analytical model, consisting of  a binary system embedded in a uniform ellipsoidal density distribution. The major axis of the ellipsoid lags behind that of the binary by a constant angle, while the axis of rotation of the ellipsoid  coincides with that of the binary. The gravitational potential of an uniform  ellipsoid is given by:
\begin{equation}
\Phi(x,y,z) = \pi G \rho \, (\alpha_{0} x^{2} + \beta_{0} y^{2} + \gamma_{0} z^{2} + \chi_{0}),
\label{potential}
\end{equation}
where $\alpha_{0}$, $\beta_{0}$, $\gamma_{0}$, $\chi_{0}$ are constants given in Lamb (1879) and depend on the ratio between the principal axes of the ellipsoid. 

The treatment of the problem is an extension of that given in Goldstein (1950) for the two body problem and is analogous to the development by Boss (1984) in the context of star formation. The binary system can be treated as an equivalent one body problem,  subject to an external gravitational potential (Eq $\ref{potential}$). The Lagrangian of the system, in cylindrical coordinates, has the form
\begin{equation}
L = \frac{1}{2} \mu \left[ \left(\frac{dr}{dt}\right)^{2} + r^{2}\left(\frac{d\phi}{dt}\right)^{2} + \left(\frac{dz}{dt}\right)^{2} \right] + \frac{Gm^{2}}{r} - \frac{\pi}{2} G\rho m (\alpha_{0} r^{2} {\rm cos}^{2}(\Delta\phi) + \beta_{0} r^{2} {\rm sin}^{2}(\Delta\phi) + \gamma_{0} z^{2} + \chi_{0}),
\end{equation}
where $\Delta\phi$ is the angle between the major axis of the ellipsoid and the axis between the binary members. As expected from the self similar behavior, in our simulation (\S 4.1) we found that $\Delta\phi$ is approximately constant. In this simple model we assume that the angle $\Delta\phi$ is  constant and this produces an explicit dependence of the Lagrangian on the coordinate $\phi$, therefore its canonical momentum is not conserved. The Euler-Lagrange equation for the coordinate $\phi$ is
\begin{equation}
\frac{d}{dt}\left(r^{2}\frac{d\phi}{dt} \right) = -2\pi G (\beta_{0} -
\alpha_{0}) \, {\rm cos}(\Delta\phi) \, {\rm sin}(\Delta\phi) \, \rho \, r^{2}. 
\label{eqphi}
\end{equation}

\begin{figure}
\plotone{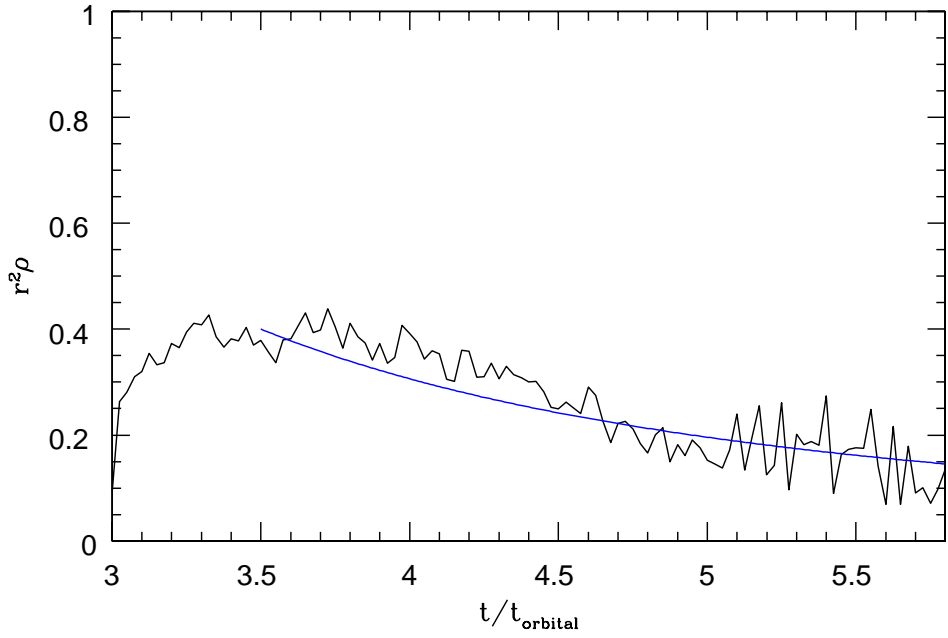}
\caption{The plot shows the evolution of the product ${\rm r}^{2} \rho$, where r is the binary searation and $\rho$ the mean density of the ellipsoid. The black line is the result of the SPH calculation and the blue line represents the function ${\rm 4.9 \cdot (t/t_{orbital})^{-2}}$
\label{fig12a}}
\end{figure}

\begin{figure}
\plotone{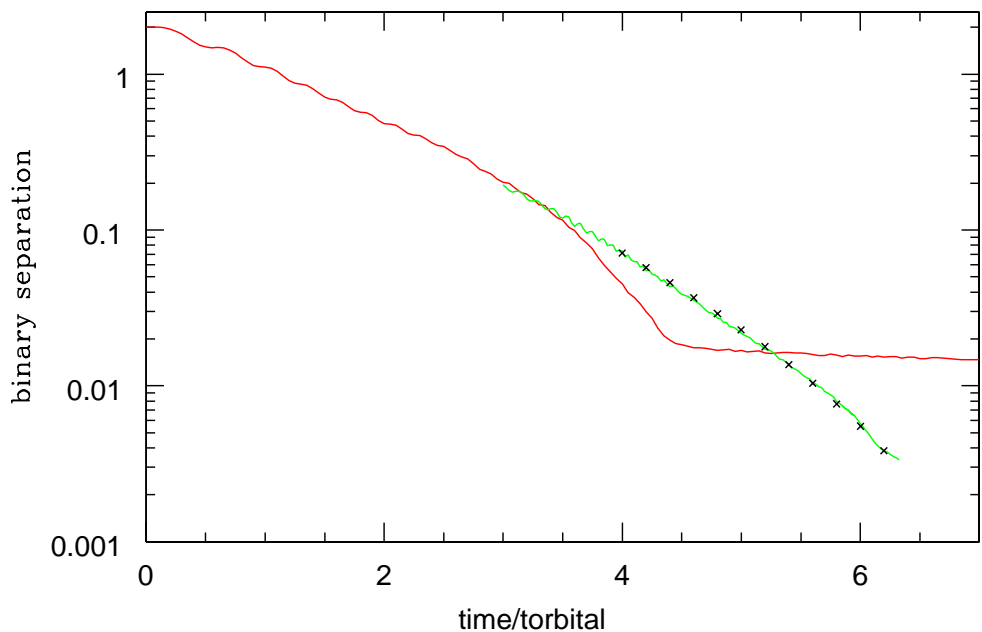}
\caption{The plot shows  the complete evolution of the binary
separation on a logarithmic scale. The red curve is the  coarse
calculation  for $M_{\rm binary}=0.02M_{\rm gas}\,$, and the green curve is the MBH's
separation in the high-resolution calculation. The black crosses show the prediction of our
simple model for the ellipsoidal torque, as described in section \S 4.2; although  simple,  it
is in  a remarkable agreement with the SPH calculation. 
\label{fig12b}}
\end{figure}

Throughout  our simulation, the mean density of the ellipsoid in the vicinity of the binary continues to  increase as the binary separation decreases, owing to the gravitational attraction of the black holes. On the right side of equation $(\ref{eqphi})$ the term that evolves with time is $\rho r^{2}$, which in our simulation varies roughly as ${\rm t}^{-2}$, as is seen in  Fig. $\ref{fig12a}$. The product $\rho r^{2}$ decreases only weakly with time because $\rho$ increases strongly as r decreases, increasing almost as $r^{-2}$. Also in this later stage, the motions of the MBHs are  almost circular, i.e.: $v \, = \, \left( (\frac{dr}{dt})^{2} + (r \frac{d\phi}{dt})^{2}  \right)^{1/2} \sim  r \frac{d\phi}{dt}$ and the circular velocity is $ v  =  \left( \frac{G M_{{\rm 12}}}{r} \right)^{1/2}$. Under those conditions  Eq $(\ref{eqphi})$ has the following solution
\begin{equation}
r(t) =   \left[r_{0}^{1/2} \,\, + \,\,  2\pi  (\beta_{0} -
  \alpha_{0}) \, {\rm cos}(\Delta\phi) \, {\rm sin}(\Delta\phi)  \,  \rho_{0} \, r^{2}_{0} \,
  t^{2}_{0} \, \sqrt{\frac{G}{M_{12}}} \,
  \left(\frac{1}{t} - \frac{1}{t_{0}}\right)\right]^{2},  
\label{rsol}
\end{equation}
where  $r_{0}$ \& $\rho_{0}$ are  respectively  the
distance  and mean density at
the time ${\rm t}_{0} = 4t_{\rm orbital}$ and $M_{12}=M_{\rm BH1} + M_{\rm BH2}$. We apply this model to a spheroid with principal axes in the ratios 2:1 (i.e. a=2b=2c). In such a configuration $\beta_{0}$=0.825 and $\alpha_{0}$=0.35. The mean angle  between the binary and the major axis  $(\Delta\phi)$ is $22.5^{o}$ and the initial mean density of the ellipsoid $(\rho_{0})$ is
10. The black crosses in Fig. $\ref{fig12b}$ represents the  separation given by Eq $(\ref{rsol})$, and  by comparing with the SPH run (green line), we see that this simple model describes successfully the evolution of the binary MBH with 10\%  accuracy. Thus, we can interpret the SPH result shown in Fig. $(\ref{fig9b})$ as the angular momentum loss due to the torque produced by the ellipsoid.

\section{Application to a Galaxy Center}

We are interested in studying the evolution of a binary
MBH after a recent merger of gas-rich galaxies. During the merger
gravitational torques remove angular momentum from shocked galactic
gas causing  most of the gas to fall into the center (Barnes \&
Hernquist 1991, 1996). Those nuclear inflows fuel starburst activity
manifested by anomalous colors (Larson \& Tinsley 1978) and
enormous infrared luminosities (Sanders \& Mirabel 1996)
characteristic of  merging galaxies. Numerical simulations show that
typically  60\% of the gas from the progenitor galaxies ends up in a
nuclear cloud with dimensions of several hundred pc (Barnes \& Hernquist
1991, 1996). On these scales, the galactic gas dominates the
gravitational potential. The best examples  of such configurations are
the Ultra Luminous Infrared Galaxies (ULIRG), which typically have
$10^{10} \, \msun$ of  atomic and molecular gas in
their  inner regions (r $\leq$ few 100pc). Our model can be scaled to
represent the conditions in the inner regions of an ULIRG, the time
conversion of the initial orbital period into physical units being  

\begin{equation}
{\rm t_{orbital}} =  \left( \frac{L}{100 \rm pc} \right)^{3/2} \, \left( \frac{2.5
    \cdot 10^{10} \Msun}{M_{\rm gas}} \right)^{1/2} \,\, 7.4 \cdot
    10^{5} \, {\rm yr} 
\end{equation}
 
 
%

Thus, if we convert the simulation with $M_{\rm BH}$=0.01 $M_{\rm gas}$  into
 physical units,  a  binary with $M_{\rm BH} = 10^{8} \msun$ will
reduce its separation from 200 pc to 0.1 pc in only  5.5 $\cdot 10^{6}$ yrs. 

We don't find any sign  of  ejection of the surrounding
gas, as happens with the stars in the later stages in the evolution of a binary MBH
 in a stellar system (Begelman, Blandford \& Rees 1980; Makino \&
Ebisuzaki 1996; Milosavljevic \& Merritt 2001,2003). Therefore  we expect
that the torque mechanism that we have described will continue to operate until additional physical effects become relevant (e.g. radiation pressure). In any case, at the end of our
simulation, the black hole binary  is entering the regime where
gravitational radiation is efficient (see \S 5.1), driving the final
coalescence of the MBH. Therefore it is reasonable to expect that the binary MBH will coalesce in a merger of gas-rich galaxies.

\subsection{Gravitational Radiation}

The final phase in the evolution of a binary MBH occurs when eventually the MBHs  become close enough to allow  gravitational radiation to become an efficient mechanism for angular momentum loss.

The gravitational radiation timescale (Peters 1964) is 
\begin{equation}
t_{\rm gr} = \left(\frac{\dot{a}}{a}\right)^{-1} = \frac{5}{64} \, \frac{c^{5}a^{4}}{G^{3}M_{12}^{3}} F(e)
\end{equation}
where $M_{12}=M_{\rm BH1} + M_{\rm BH2}$, $a$ is the binary separation and $F(e)$ is:
\begin{equation}
F(e) = \frac{(1-e^{2})^{7/2}}{1+\frac{73}{24}e^{2}+\frac{37}{96}e^{4}}
\end{equation}
$F(e)$ contains the eccentricity dependence which is a weak  for small
e; e.g. $F$(0)=1 and $F$(0.5)$\sim$0.205.
  
At the end of our simulations we reach a final separation of 0.1
pc. At this distance the gravitational radiation timescale is ${\rm t}_{\rm gr} \sim 1.2\cdot 10^{8}$yr. We feel confident that we can  extrapolate our results to  smaller separations, because the ellipsoidal torque model describes successfully the evolution of the binary MBH. If we extrapolate our results  to a separation of 0.05pc, we predict that
the binary MBH will merge in only $\sim 10^{7}$yr. This timescale is 
smaller than the time
that star formation takes to consume most of the gas in  such a dense
cloud (a few $10^{8}$ yr), justifiying our assumption that the binary interacts with a gaseous background instead of with a stellar one. 

\subsection{Another Application: Formation of Close Binary Systems}

An important problem in  star formation theory is that of the formation of close binary stars. Orbital decay  of forming binary systems due to  gravitational drag may help to explain the distribution of binary separations, since all of the proposed formation mechanisms (fission, fragmentation, capture, disintegration of a larger stellar system) predict an insufficient number of close binary systems (see Bonnell 2001 for a review). Bate, Bonnell \& Bromm (2002) show that gravitational interaction of a binary with its surrounding medium is a possible mechanism to form  close binary systems. Our  work  has enough resolution to study such  systems in more detail than Bate, Bonnell \& Bromm (2002).
 
We scale our model to represent the conditions in star forming regions. Thus, the time conversion of the initial orbital period into physical units is
\begin{equation}
{\rm t_{orbital}} = \left(\frac{\rm M_{gas}}{\rm 10 \, M_{\sun}}\right)^{-1/2} \, \left(\frac{\rm D}{\rm 0.01 \, pc}\right)^{3/2} \, 3.6 \times 10^{4} \, {\rm yr} \, ,
\end{equation}
where $M_{gas}$ is the total mass of the gas and D is the distance unit. 

We apply our result to  the case of low mass star formation, assuming a total gas mass  of 10$M_{\sun}$ and an initial separation between the protostars of 0.02pc. This gives an initial orbital period  of $3.6 \times 10^{4}$ yr. Thus we predict that a binary with 0.1$M_{\sun}$ protostars will reduce its separation from 0.02 pc to 2 AU in 7 initial orbital periods (Fig. $\ref{fig9b}$), or only $2.6 \times 10^{5}$ years.

Since the gravitational drag timescales are  predicted to be comparable with those of the star formation process, gravitational drag can play  an important role  in the formation of stars and can determine the final separations of binaries. Therefore, the orbital evolution of binary systems due to gravitational drag is a promising mechanism for the formation of close binary stars.

\section{Conclusions}

In this paper we study the role of gas in driving the evolution of a binary MBH, and we follow its evolution through many orbits and close to the point where gravitational radiation becomes important. We present results for a relatively idealized  case in which the gas is assumed to be in a nearly spherical and relatively smooth distribution.

There are important   differences in the long term
evolution of a binary MBH in a gaseous medium,  compared to a stellar background. From N-body simulations,  it is expected that the merging of an MBH 
binary eventually stalls  in a  stellar background. However, our simulations
suggest that the MBH binary will finally merge if it's  embedded in a
gaseous medium.

In the early evolution of  a binary MBH, the separation diminishes
due to the gravitational drag exerted
by the background gas. This decay is faster than in a stellar background by a
factor $\sim$1.5 for a given density profile. 
In the later stages, when  the binary MBH dominates the gravitational
potential in its vicinity, the medium responds forming 
 an ellipsoidal configuration.  The axis of the ellipsoid lags behind  the binary axis, and this  offset  produces a torque on the binary that is now  responsible for the continuing loss of angular momentum. The evolution of the system is approximately self-similar because the size of the ellipsoid shrinks with  the binary separation. This torque is able to
reduce the binary separation to distances where  gravitational radiation is
efficient. We don't find any sign of ejection of the surrounding gas, as happens with stars in the later stages in the evolution of a binary MBH in a stellar system. Moreover, the gas density $\rho$ in the vicinity of the binary  increases strongly as the separation r decreases, increasing almost as $r^{-2}$. 

Assuming typical parameters from observations of ULIRG, the overall binary evolution has a merger timescale of $10^{7}$ yrs. This timescale is fast enough to support the
assumption that the binary is in a mostly gaseous background medium,
because the timescale for gas depletion in a starburst region is typically  several times $10^{7}$
yrs. Galaxies typically merge in  $10^{8}$ yrs, and therefore these results imply that in a merger of gas-rich galaxies the BHs will coalesce
soon after the galaxies merge. The final coalescence has crucial implications for possible scenarios of massive black hole evolution and
growth. In particular, this result supports scenarios where hierarchical build-up of massive black holes plays an important role. The final coalescence of the black holes lead to gravitational radiation emission that would be detectable up to high redshift by LISA.

In subsequent papers, we will present results of calculations in progress for
 cases in which the gas is cooler, and has a more flattened and
clumpy spatial distribution. Preliminary results show again, as in the case of smooth gas, a large decrease of the binary separation in a few initial orbital periods, and thus support the more general applicability of the results presented here for an idealized case. 

\section{Acknowledgments}

A. E. thanks Fundaci\'on Andes  for fellowship support. We are grateful to Volker Springel for making available to us a version of GADGET and we thank Priya Natarajan for valuable comments. Some of the simulations were performed at the Universidad de Chile computer cluster, funded by FONDAP grant 11980002.

\appendix


\section{Code Testing}

In order to validate the proper use of GADGET, we have done additional
tests to those   performed by the code authors  (Springel,
Yoshida \& White 2001).

\subsection{Effects of Numerical Diffusion}

One of the disadvantages of using multidimensional Lagrangian schemes is
that they suffer from numerical diffusion, introduced by the appearance of
vortices that requires remapping the tangled and twisted code grid.  The SPH
technique can be viewed as an intrinsic remapped Lagrangian grid which 
implies a sizable numerical diffusion.
 
\begin{figure}
\plotone{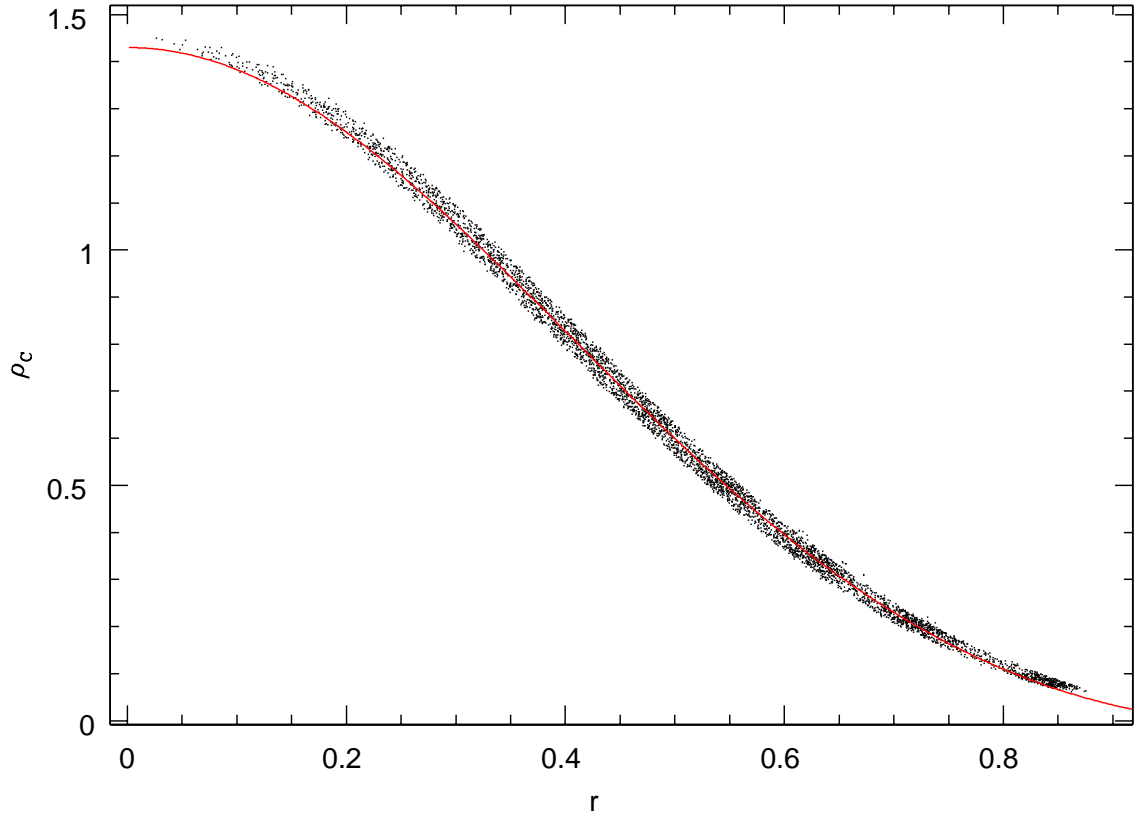}
\caption{Density profile of an n=1.5 polytrope. The points are the density values for the 5000 SPH particles in the final relaxed configuration. The solid line is the solution of the Lane-Emden equation. Good agreement is found between the exact solution and the SPH prediction. 
\label{fig1}}
\end{figure}

To test the effects of numerical diffusion in the GADGET code, we
simulate the adiabatic oscillation of a polytrope  following
Steinmetz \& Muller (1993) and investigate the damping due to numerical diffusion.
We first need to establish the hydrostatic equilibrium configuration, which is also
a test of the interaction between the gravitational and the hydrodynamic
elements of the code.  We chose a density profile given by the solution to
the Lane-Emden equation for n=3/2, and let it relax under the polytropic
equation of state
\begin{equation}
p = K \, \rho^{5/3}.
\end{equation}
Fig. $\ref{fig1}$ compares the resulting density distribution from the SPH
code using 5000 particles with the Lane-Emden solution (Kippenham and Weigert
1991).   The dots follow the curve throughout the cloud radius to
within a 5\% of spread in density.

\begin{figure}
\plotone{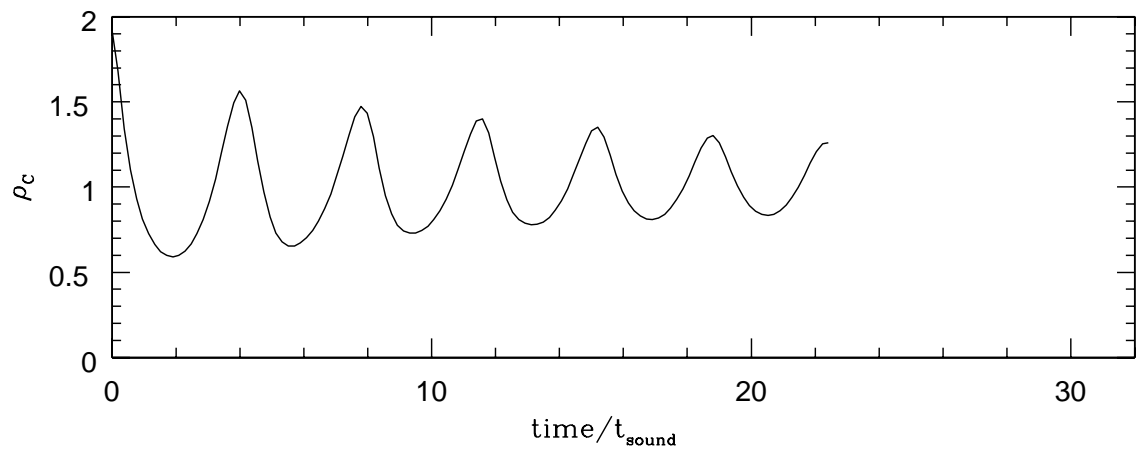}
\caption{Pulsating polytrope. The plot shows the evolution of the central density through several sound crossing times. The oscillation frequency $(\nu_{osc} \sim 1/(4t_{\rm sound}))$ is correctly predicted, but the pulsation amplitude decreases noticeably due to the inherent numerical diffusion of the SPH technique.
\label{fig2}}
\end{figure}

In order to test the damping of an adiabatic oscillation due
to numerical diffusion, we imposed (on the equilibrium model) a
spherically homologous contraction by a factor 0.8 in radius in order
to produce strong pulsations of the polytropic  sphere. Fig.
$\ref{fig2}$ shows the evolution of  the central density as a function
of time, and the oscillation frequency has the  predicted value
$\nu_{\rm osc} \sim 1/(4t_{\rm sound})$. 

The pulsation amplitude decreases in time  due to the numerical
diffusion of the code. After 5 pulsation periods it has decreased by a
factor 2.5. This diffusion depends on the artificial
viscosity used, in this case (shear-reduced artificial viscosity)
the results are similar to that found by Bate (1995). 
Other prescriptions for artificial viscosity produce less
damping in the oscillation, but those are not suitable for our problem
because they produce a higher loss of angular momentum due to the extra shear viscosity.


\subsection{Shock Modeling}

The supersonic motion of MBHs in a gaseous medium will produce strong
shocks, and therefore it is crucial to know if the code is able to resolve
shocks.  We test the ability of GADGET to handle shocks by following the
adiabatic collapse of an initially isothermal gas sphere, which is a common
test case for SPH codes (Evrard 1988, Hernquist \& Katz 1989, Steinmetz \&
Muller 1993, Thacker et al. 2000).  The initial (non equilibrium) state consists of an
adiabatic gas sphere with total mass M, radius R and density profile
\begin{equation}
\rho(r) = \frac{M}{2\pi R^{2}} \, \frac{1}{r} 
\end{equation}
The gas is initially isothermal, with a specific internal energy of
\begin{equation}
u = 0.05 \, \frac{GM}{R}
\end{equation}

The thermal energy is not enough to provide pressure support, therefore the gas
sphere begins to collapse.  As the collapse proceeds the gas is heated until
the core temperature rises enough to produce a central bounce and a shock
wave is formed and propagates outwards (Evrard 1988).  After the
shock has passed through most of the gas, the gas sphere should reach
virial equilibrium. The evolution of the kinetic, thermal, gravitational and total energies have
been widely discussed in the literature (Evrard 1988, Hernquist \& Katz
1989, Steinmetz \& Muller 1993, Thacker et al. 2000) and the results for the
GADGET code are published in Springel, Yoshida \& White (2001).  We focus
our attention on the outward propagating shock and test how well is handled.

\begin{figure}
\plotone{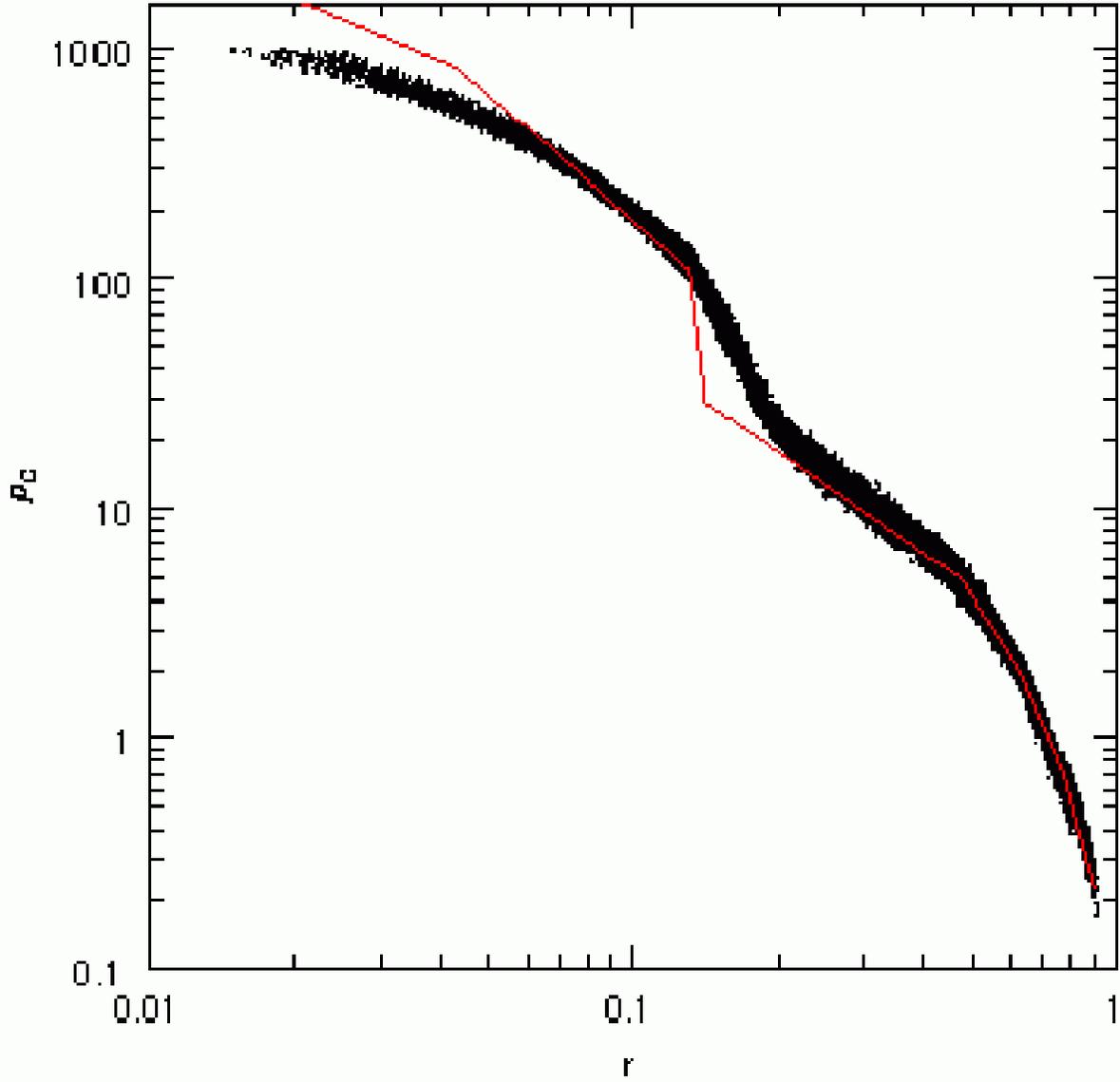}
\caption{Density profile of the collapsing  sphere at t=0.8; the outward propagating shock is clearly seen between r=0.15 and r=0.25. The red curve is the result obtained by a finite difference method.
\label{fig3}}
\end{figure}

Fig. $\ref{fig3}$ shows the density  of the collapsing system obtained for a
simulation with 30,000 particles at t=0.8.  Initially the gas sphere is at rest.  We
obtain the coordinates from a distorted regular grid that reproduces the
density profile and we give our results in units of M = R = G = 1.  The
shock propagating outward is clearly visible  between r=0.15 and
r=0.25 in the figure. The results agree very well with the  results obtained by a finite difference method (Thomas 1987; red solid
line).  The results from Fig. $\ref{fig3}$ are also in very good agreement with
those found by Hernquist \& Katz (1989),  Steinmetz \& Muller
(1993) and even with the results obtained by the best artificial viscosity implementation of Thacker et al. (2000).

\subsection{Collapse of rotating sphere}

The collapse of a rotating sphere has become a standard test for galaxy
formation codes (Navarro \& White 1993).  One relevant
application of our work is to study the formation of MBH in
protogalaxies. The spherical 'protogalaxy' is composed of dark 
matter providing 90\% of
the mass and baryons providing the remaining 10\%.  Initially the
cloud has a density profile of $\rho(r) \propto r^{-1}$ .  The velocities are chosen so that the sphere
is in  solid rotation with a spin parameter 
\begin{equation}
\lambda = \frac{L E^{1/2}}{G M^{5/2}} \, \sim \, 0.1
\end{equation}
the initial radius of the cloud is $100\Kpc$, and the total mass (gas and
dark matter) is $10^{12}\Msun.$ The dynamical time-scale of
the system is then $\tau_{\rm dyn} \sim 3.7 \times 10^{8}$yr. The gas is
initially at a uniform temperature of $T=10^{3}K$ that is well below the equilibrium virial
temperature of the system ($T_{\rm virial} \sim 10^{6}$K). We use the
same units as  Navarro \&
White (1993): G=1, $\rm [mass]=10^{10}\Msun$, $\rm [time]=4.71 \,
10^{6}\rm yr$, $\rm [distance]=1\Kpc.$

\begin{figure}
\plotone{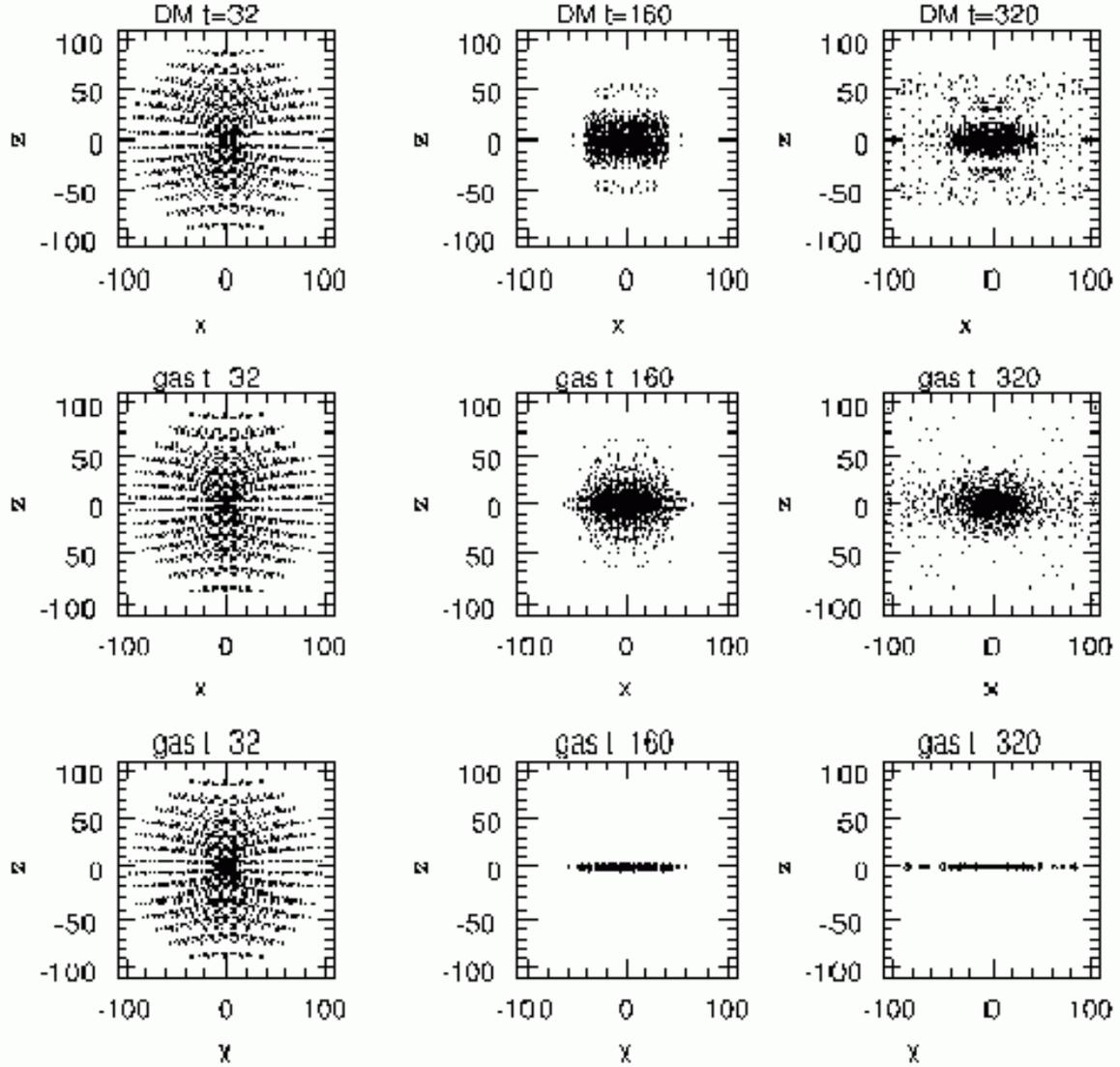}
\caption{Collapse of a primordial rotating cloud. a) The evolution of
  the dark matter halo in the X-Z plane, in three given times. b) The 
 evolution of the  gaseous component in the X-Z plane, without any
  radiative cooling. c) The evolution of the gaseous component in the
  X-Z pane, with primordial radiative cooling.
\label{fig4}}
\end{figure}

As  in the case of Evrard collapse (A.2), the gas evolution is
dominated by a shock wave that propagates outward and transforms most of the
kinetic energy into heat.  Fig. $\ref{fig4}$(a,b) shows the evolution
of the gas and the dark matter components, at three given times: t=32,
160, 320. The dark matter is
virialized soon after collapse forming a tight core (Fig. $\ref{fig4}$a). The gas shows a distribution similar to the
dominant dark matter except for being slightly more spherical.  The results are
consistent with those of Navarro \& White (1993).

We repeat the simulation including radiative cooling; we use a cooling
function for primordial gas interpolated from Sutherland \& Dopita (1993).
When radiative cooling is included the evolution of the system changes
considerably because the cooling timescale in the denser regions  becomes
much smaller than the dynamical timescale of the gas; This produces radiation
losses that dissipate the thermal energy gained through shocks.  With no
pressure support, the gas collapses until it is centrifugally supported
forming a flat disk structure, as is clearly seen in Fig. $\ref{fig4}$c.

\section{Particle Splitting}
    
Particle splitting   (Bromm 2000, Kitsionas 2000, Bromm \& Loeb 2003) is a  procedure designed to increase the resolution of a SPH code in some regions where it
is required. In this procedure, all  particles in the region of
interest are  replaced by  set of smaller particles, increasing the
resolution of the simulations. We refer to the smaller particles as
'fine' particles and the original ones as 'coarse' particles. 

In our simulations we need to increase the resolution in the later
stages, to be able to resolve the region inside r = 0.1, where
 most of the relevant dynamics occur. We stopped the simulation just before
the MBHs arrive to the inner region (r $<$ 0.1). We then resampled the fluid by splitting each SPH particle from the original simulation (called 'parent' particle and denoted by p) into $N_{\rm split}$ particles (called 'child' particles and denoted by c). 

The child particles are randomly distributed according to the SPH
smoothing kernel W$(r_{\rm c}-r_{\rm p};h_{\rm p})$ where $h_{\rm p}$
is the smoothing length of the parent particle (see Springel, Yoshida \& White 2001 for the kernel implementation used by GADGET). The velocity is directly inherited from the parent particle, i.e. $\vec{v_{\rm c}}=\vec{v_{\rm p}}$ and each child particle is assigned a mass of $m_{\rm c}$ = $m_{\rm p}/N_{\rm split}.$      

We apply the resampling procedure to the whole sphere instead of only
the inner region. The reason is that in the boundary between the
coarse and fine regions the density is overestimated for the coarse
particles and underestimated for the fine ones. The net result is an
inter-penetration of the coarse particles into the r $<$ 0.1 region,
when they become close enough to the MBH to produce  a numerical
overestimate of the drag force. Kitsionas (2000) tried to solve this
artifact by changing the SPH formalism from  a fixed total number of
neighbors to a fixed total mass of the neighbors. This reduced the
overestimate of the fluid properties for the coarse particles but was
not able to prevent the  penetration of coarse particles into the fine
region. For that reason we choose to resample the whole fluid although
it is numerically more expensive.


\end{document}